%% file: main.tex
\documentclass[conference,9pt]{IEEEtran}

\usepackage{amsmath,amssymb}
\usepackage{graphicx}
\usepackage[table]{xcolor}
\usepackage{tikz}
\usetikzlibrary{arrows.meta,positioning,calc}
\usepackage{booktabs}
\usepackage{multirow}
\usepackage[ruled,vlined]{algorithm2e}
\usepackage{tabularx}
\usepackage{cite}
\usepackage{pifont}
\usepackage{hyperref}
\usepackage{makecell}

\ifCLASSOPTIONcompsoc
\usepackage[caption=false,font=normalsize,labelfont=sf,textfont=sf]{subfig}
\else
\usepackage[caption=false,font=footnotesize]{subfig}
\fi

\definecolor{avgcol}{HTML}{FDEBC8}
\definecolor{alacol}{HTML}{E8F0FE}
\definecolor{avghi}{HTML}{FAC559}
\definecolor{alahi}{HTML}{A9C7F8}
\definecolor{qwrong}{HTML}{B23A2E}
\definecolor{qright}{HTML}{1B7A43}
\definecolor{spanrow}{HTML}{EAF5EE}
\definecolor{inputrow}{HTML}{ECECEC}

\definecolor{qualgt}{HTML}{E3A52B}
\definecolor{qualadv}{HTML}{6E92D6}
\newcommand{\qgt}[1]{\textbf{\textcolor{qualgt}{#1}}}
\newcommand{\qadv}[1]{\textbf{\textcolor{qualadv}{#1}}}
\newcommand{\qck}{\ding{51}}
\newcommand{\qx}{\ding{55}}
\newcolumntype{Y}{>{\raggedright\arraybackslash}X}

\newcommand{\inputcell}[3]{%
  Spoken content: ``#1''\\
  Voice attribute: \qck~\qgt{#2}\\
  Adversarial label: \qx~\qadv{#3}.%
}

\newcommand{\method}{IAAN}
\newcommand{\eg}{e.g.,}

\newcommand{\up}{$\uparrow$}
\newcommand{\dn}{$\downarrow$}

\begin{document}
\flushbottom

\title{Enhancing How Large Audio-Language Models Hear: Identifying and Amplifying Acoustic Neurons\\ in the Audio Encoder}

\author{ \IEEEauthorblockN{Yu-Han Huang\textsuperscript{1,4}, Chih-Kai Yang\textsuperscript{2,*}, Ke-Han Lu\textsuperscript{2,*}, An-Yu Cheng\textsuperscript{1,*}, Hung-yi Lee\textsuperscript{3}} \IEEEauthorblockA{ \textsuperscript{1}Department of Electrical Engineering, National Taiwan University, Taiwan \\ \textsuperscript{2}Graduate Institute of Communication Engineering, National Taiwan University, Taiwan \\ \textsuperscript{3}NTU Artificial Intelligence Center of Research Excellence (NTU AI-CoRE), Taiwan\\ \textsuperscript{4}ASUS Open Cloud Infrastructure Software Center, Taipei, Taiwan\\ } } \maketitle \let\thefootnote\relax\footnotetext{\textsuperscript{*}Equal contribution.}

\input{sections/00_abstract_v2}

\input{sections/01_introduction_v2}
\input{sections/02_related_work_v2}

\input{sections/03_method_v2}
\input{sections/04_experimental_setup_v2}

\input{sections/05_results_v2}

\input{sections/06_analysis_v2}

\input{sections/08_conclusion}

\section*{AI-Generated Content Disclosure}
AI tools were used only for language editing; the authors reviewed all content and take full responsibility.

\section*{Acknowledgment}
This work was supported by the Ministry of Education (MOE) of Taiwan under the project Taiwan Centers of Excellence in Artificial Intelligence, through the NTU Artificial Intelligence Center of Research Excellence (NTU AI-CoRE). Also, we would like to thank
Chung-Cheng Chen, Jen-Hao Cheng, TSUNG-YING YANG, Hsiao-Tsung Hung, Dau-Cheng Lyu
at ASUS-OCIS for their helpful feedback and discussion throughout this work.

\bibliographystyle{IEEEtran}
\bibliography{refs}

\end{document}

%% file: sections/00_abstract_v2.tex
\begin{abstract}

Large audio-language models (LALMs) often underperform on fine-grained, non-semantic attributes of speech, such as a speaker's emotion, despite strong performance on speech content. Improving this without the cost of retraining calls for an effective inference-time intervention, yet most existing methods intervene only after the audio encoder and operate at a relatively coarse granularity. The encoder itself, where acoustic information is first extracted from the waveform, remains largely unexplored, especially at the level of individual neurons. To this end, we introduce IAAN, Identifying and Amplifying Acoustic Neurons, a training-free and label-free method that scores each feed-forward neuron in the audio encoder by contrasting its activation on the real waveform with that on a noise reference lacking the real audio's acoustic information. IAAN then amplifies a small set of the highest-scoring neurons at inference. Across ten non-semantic speech attributes, IAAN improves average accuracy by 25.7 points on Audio-Flamingo-3, 21.4 on Qwen2.5-Omni, and 9.7 on Kimi-Audio. It also continues to improve a model that has already been explicitly fine-tuned to prioritize acoustic evidence. In controlled comparisons, both the encoder locus and neuron-level selectivity prove necessary for this gain. Intervening after the encoder, at the decoding side or inside the language model, yields little to no improvement, or even deteriorates accuracy. The improvement also depends on which specific neurons are amplified, not merely on their number, confirming that IAAN's acoustic score succeeds in identifying the neurons that matter. These results show that a small, precisely targeted intervention inside the audio encoder is an effective and largely untapped way to strengthen the acoustic understanding of LALMs, opening a new direction for inference-time methods that improve acoustic perception through neuron-level access to the encoder.
\end{abstract}

\begin{IEEEkeywords}

large audio-language models, paralinguistic understanding, audio encoder, inference-time intervention, neuron activation steering
\end{IEEEkeywords}

%% file: sections/01_introduction_v2.tex
\section{Introduction}

Large audio-language models (LALMs)~\cite{audioflamingo, audioflamingo2, audioflamingo3, qwenaudio, qwen2audio, qwen25omni, desta, desta2, desta25, salmonn, lin2025preliminary, yang2024building, kimiaudio} build on a pretrained large language model (LLM) backbone with an audio encoder~\cite{whisper, beats}, extending auditory understanding~\cite{audiolens, yang2025sake, akb} to the LLM. Speech carries rich information beyond the words themselves, such as a speaker's emotion or gender, yet current LALMs handle semantic content well while performing markedly worse on such fine-grained, non-semantic attributes~\cite{audio-eval-survey, mmau, voxparadox, speechcopilot, resurfacing, audiobench, sakura, voxprofile, echomind}. Our goal is to strengthen this weaker side, making LALMs perceive acoustic detail in the voice more reliably without retraining.

Most inference-time efforts to close this gap intervene on the language model backbone of LALMs or in its decoding process, through contrastive decoding over output logits~\cite{aad, lin2026contrastive, apscd} or steering of the backbone's hidden states, attention, or neurons~\cite{avs, nudginghidden, audiolens, audiospecialistheads, neuronemotion, textbias}. Because these operate after the audio representations reach the backbone, their effectiveness is bounded by how much acoustic evidence those representations still carry~\cite{modalitygap, mohebbi2024disentangling}. The audio encoder, where that evidence first arises~\cite{whisperat, pasad, whisperemointerp}, has received comparatively little attention as a site of intervention. A separate question is the granularity at which such an intervention should act. Output logits offer only a coarse handle on the model's overall behavior, and hidden-state interventions, though finer, still act on a whole layer at once. Individual feed-forward neurons are finer still, raising the question of
whether some neurons inside the audio encoder respond more strongly to acoustic information than others. \emph{Where to intervene, and at what
granularity, thus remain open.}

Motivated by these open questions, we introduce IAAN (Identifying and Amplifying Acoustic Neurons), a training-free and label-free method that intervenes inside the audio encoder at the level of individual neurons (Fig.~\ref{fig:overview}). Our intuition is simple: a neuron encoding a genuine acoustic attribute should respond more strongly to real audio than to a reference clip of unstructured noise, which carries no such attribute. IAAN scores each encoder neuron by contrasting its activation on the real audio with that on the noise reference, then amplifies the highest-scoring ones during inference. Since the audio encoder ($\sim$0.6B) is far smaller than the LLM backbone ($\sim$7B), these additional encoder passes add little overhead.

We evaluate IAAN on three open-source LALMs, Audio-Flamingo-3, Qwen2.5-Omni,
and Kimi-Audio, using VoxParadox~\cite{voxparadox}, a benchmark of ten
non-semantic speech attributes spanning speaker traits, prosodic cues, and
signal-level properties. IAAN improves average accuracy on all three models by up to +25.7 points
(13.1\% to 38.8\% on Audio-Flamingo-3), with consistent gains of +21.4 and
+9.7 points on Qwen2.5-Omni and Kimi-Audio.
Controlled comparisons confirm that both the encoder locus and neuron-level
selectivity are necessary for this gain. Applying analogous interventions
after the encoder, at the decoding side or inside the language model, changes
the model's predictions while yields little to no accuracy gain. Within the
encoder, the gain depends on which neurons are amplified rather than on scale
or location alone. The effect also carries over
to free-form descriptions, which after IAAN more often report the acoustic
attributes actually present in the voice.

Overall, our contributions are three-fold:
\begin{itemize}

\item We identify individual feed-forward neurons in the audio encoder as an effective and largely overlooked site for improving acoustic
perception in LALMs, in both locus and granularity.

\item We propose IAAN, a training-free intervention that scores encoder neurons by contrasting their responses to real audio and to a reference input, then amplifies the  subset in audio encoder.

\item We show across three LALMs that IAAN substantially improves accuracy on a broad suite of non-semantic auditory tasks, and through controlled ablations establish that both the encoder locus and the selectivity of the neuron set are necessary for the gain.

\end{itemize}

\input{figs/fig_overview}

%% file: figs/fig_overview.tex
\begin{figure}[t]
  \centering
  \includegraphics[width=\linewidth]{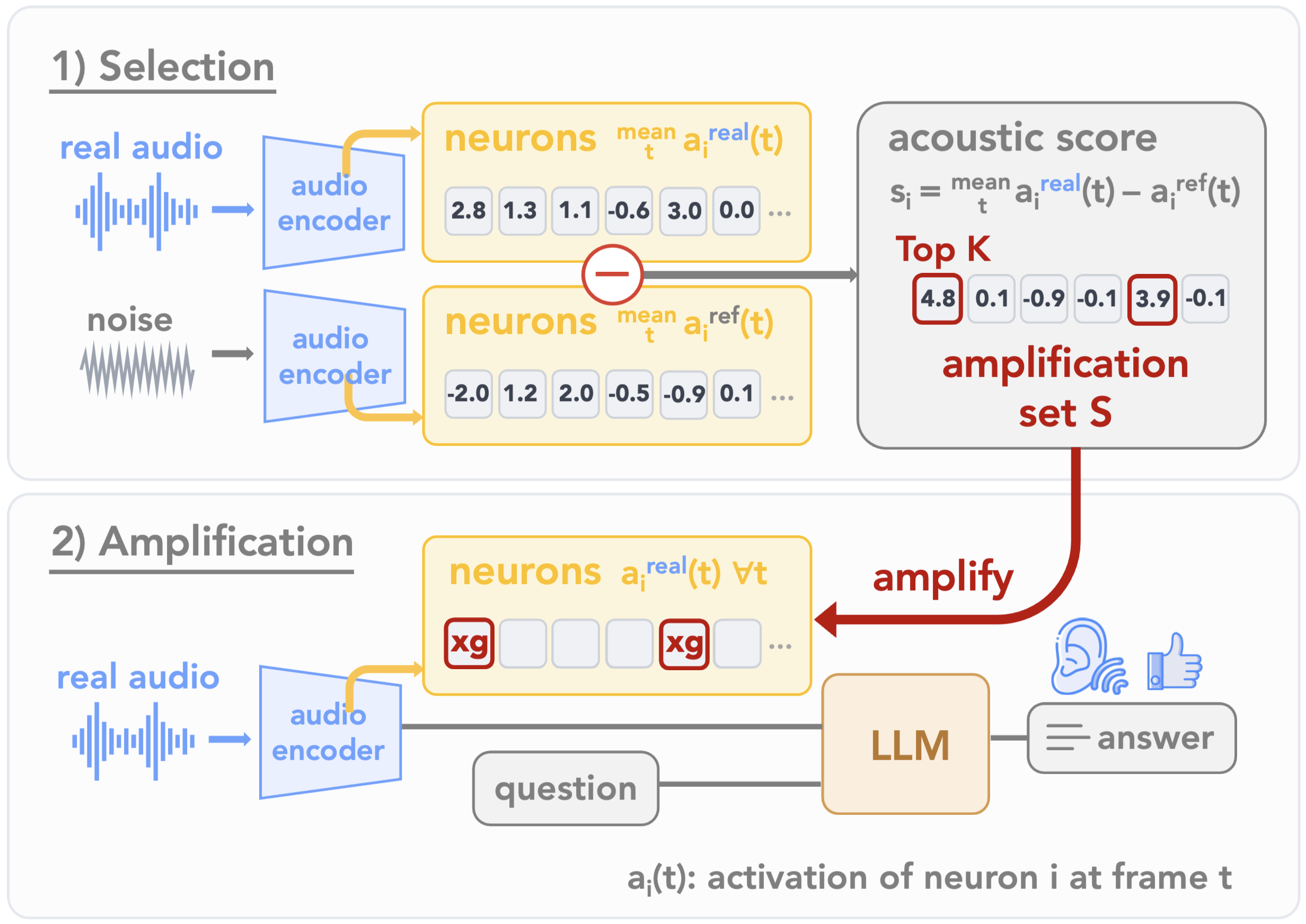}
  \caption{Overview of \method{}. The encoder first runs on the real
    audio $x$ and on a noise reference $r$ to score each feed-forward
    neuron by its acoustic score $s_i$ (Eq.~\eqref{eq:acoustic}); the
    top-$K$ neurons are then amplified during the encoder pass on the
    real audio that produces the answer.}
  \label{fig:overview}
\end{figure}

%% file: sections/02_related_work_v2.tex
\section{Related Work}
\label{sec:related}

\subsection{LALMs and Audio Understanding Benchmarks}

Large audio-language models (LALMs)~\cite{audioflamingo, audioflamingo2, audioflamingo3, qwenaudio, qwen2audio, qwen25omni, desta, desta2, desta25, salmonn, lin2025preliminary, yang2024building, kimiaudio} pair a pretrained language-model backbone with an audio encoder, extending the language model to the auditory modality. Their capabilities have been evaluated through benchmarks covering instruction following~\cite{speechifeval}, general audio comprehension~\cite{dynamicsuperb, dynamic-superb-2, airbench}, reasoning~\cite{mmau, sakura, mmar, yang2026mugen}, fairness~\cite{listenfairly}, and robustness~\cite{emotionvuln, hearingorder, textbiasmcr}. Benchmarks focusing on fine-grained acoustic attributes beyond a speech's semantic content show that current LALMs still struggle to reliably capture and use such acoustic evidence~\cite{sdeval, cpbench, lexicalacoustic, voxparadox, tau, audiobench, voxprofile, echomind}. Among these, we adopt VoxParadox~\cite{voxparadox} as our testbed, as it
directly probes fine-grained acoustic ability across ten attributes.

\subsection{Inference-Time Intervention}
Inference-time intervention aims to influence the behavior of a frozen model without retraining. In text LLMs, prior work has intervened at different levels of granularity. At the output level, contrastive decoding modifies logits by contrasting a stronger model or input condition against a weaker one~\cite{contrastivedecoding, dola}. At the representation level, activation steering adds task- or behavior-related directions to hidden states~\cite{actadd,caa,iti,repe}. At a finer granularity, neuron-level methods edit or control individual feed-forward units, motivated by evidence that such neurons can store or express specific knowledge~\cite{geva2021,knowledge_neuron,rome}.

Similar ideas appear in LALMs. Decoding-side methods contrast logits with and without audio input to make generation more audio-aware~\cite{aad, lin2026contrastive, avcd, tcd, apscd}. Hidden-state steering modifies internal representations using audio-derived directions or reasoning-oriented nudges~\cite{avs,nudginghidden,accentsteer}. Neuron-level intervention has also been explored in speech-generative LALMs for emotion control, with neurons selected using emotion labels~\cite{neuronemotion}. However, they mostly intervene after the audio encoder, either at decoding or
inside the language-model backbone, whereas the encoder is where acoustic
information is first extracted, making it a natural site for intervention.

Our method, \method{}, moves the intervention into the audio encoder, identifying and amplifying a sparse, input-specific set of feed-forward neurons whose activations are stronger on real audio than on a reference signal without meaningful acoustic cues. In contrast to prior LALM interventions, \method{} is encoder-side, neuron-level, and label-free, requiring no calibration set, attribute labels, fixed steering direction, or additional LLM decoding pass.

%% file: sections/03_method_v2.tex
\section{\method{}: Identifying and Amplifying Acoustic Neurons}
\label{sec:method}

We introduce \method{}, a training-free and label-free intervention that operates entirely within the audio encoder of LALMs at inference time, as illustrated in Fig.~\ref{fig:overview}. The audio encoder runs two forward passes, one on the real
audio waveform and one on a noise reference lacking the real audio's acoustic information, and we
contrast each feed-forward neuron's activation across the two passes to obtain a per-neuron
\emph{acoustic score} measuring how strongly it responds to acoustic information
(Sec.~\ref{ssec:acoustic}). At inference, we rank neurons by this score and amplify the
top-scoring ones (Sec.~\ref{ssec:steer}). This raises the model's reliance on the acoustic
signal in its representations, computed on the fly per clip without any training or attribute
label.

\iffalse
The remainder of this section defines the neurons we amplify
(\S\ref{ssec:neurons}), the acoustic score
(\S\ref{ssec:acoustic}), the reference signal (\S\ref{ssec:ref}), the amplification operation (\S\ref{ssec:steer}), and the controls that isolate \emph{which} property of the
selected set drives the effect (\S\ref{ssec:controls}).
\fi

\subsection{Preliminary: Feed-Forward Neurons}
\label{ssec:neurons}
A transformer block applies multi-head attention followed by a two-layer feed-forward network (FFN) to a residual-stream vector $h$: $\mathrm{FFN}(h)=W_2\phi(W_1h)$, with hidden width $d_{\text{ff}}$ and elementwise nonlinearity $\phi$. We define each coordinate $a_i = \phi(W_1 h)_i$ of the
FFN hidden activation $a=\phi(W_1h) \in \mathbb{R}^{d_{\text{ff}}}$, i.e.\ each individual input dimension
read by the second projection $W_2$, as a \emph{neuron}~\cite{knowledge_neuron, language_neuron}. Since the audio encoders are transformers, applying this definition to an encoder with $L$ layers gives $N=Ld_{\text{ff}}$ neurons (\eg{} $N=32\times5120=163{,}840$ for a Whisper-large-v3
encoder~\cite{whisper}). We write $a_i(t)$ for the activation of neuron $i$ at frame $t$.

\subsection{Acoustic Score}
\label{ssec:acoustic}
To identify neurons that respond to acoustic information, we use a reference signal $r$ that lacks the acoustic cues in the real audio $x$, so that contrasting a neuron's activations on $x$ and $r$ isolates its acoustic sensitivity. For the reference $r$ we replace the audio with Gaussian
noise that has the same length and approximately the same mean and variance as $x$, drawn with a
fixed per-clip seed; we examine alternative choices of reference in Sec.~\ref{ssec:reference_ablation}.
The reference then takes the place of the real audio in the contrast below.

Let $a_i^{\text{real}}(t)$ and $a_i^{\text{ref}}(t)$
be the activations of neuron $i$ at frame $t$ for the real waveform and its reference.
The acoustic score is
\begin{equation}
  s_i \;=\; \underset{t}{\mathrm{mean}}\, a_i^{\text{real}}(t)
        \;-\; \underset{t}{\mathrm{mean}}\, a_i^{\text{ref}}(t).
  \label{eq:acoustic}
\end{equation}
A large positive $s_i$ marks a neuron that responds more to the real audio than to its noise
reference; a large negative $s_i$ marks one that responds less. The score is computed
independently per clip from its own real and reference activations, using no external audio
dataset.

\subsection{Selection and Amplification}
\label{ssec:steer}
We rank neurons by acoustic score and collect the top-$K$, referred to as the neuron budget, into the amplification set $\mathcal{S}$.
The score is computed once per clip from the real- and reference-pass activations, and at inference we scale
the corresponding activations during the encoder pass on the real audio, whose output is then
passed to the language model to produce the answer,
\begin{equation}
  a_i \;\leftarrow\; g\, a_i, \qquad i \in \mathcal{S},
  \label{eq:steer}
\end{equation}
with gain factor $g>1$, applied by a forward hook inside the audio encoder. \method{} thus adds encoder-only forward passes for contrasting two audio inputs to compute Eq.~\eqref{eq:acoustic}, with no extra
language-model decoding. Since the encoder is an order of magnitude smaller than the
language model, IAAN is computationally efficient.

%% file: sections/04_experimental_setup_v2.tex
\section{Experimental Setup}
\label{sec:setup}

\subsection{Models and Benchmark}
We evaluate three groups of models. \textbf{Open-source LALMs}, namely
Audio-Flamingo-3 (AF3)~\cite{audioflamingo3}, Qwen2.5-Omni (Qwen2.5)~\cite{qwen25omni}, and
Kimi-Audio (Kimi)~\cite{kimiaudio}, are the primary models to which we apply
\method{}, and the analyses in Sec.~\ref{sec:results} and
Sec.~\ref{sec:analysis_v2} focus on this group. \textbf{Specialized Tuned
Model}: AF3-PD is the model released by Pang et al.~\cite{voxparadox}, fine-tuned on paralinguistic data with an enhanced audio–LLM interface and preference optimization~\cite{dpo} to strengthen acoustic grounding. We include it to test whether IAAN remains beneficial even for a model that has been explicitly trained to rely on acoustic evidence. \textbf{Proprietary Models} (GPT-4o Audio~\cite{gpt4o} and
Gemini 2.5 Flash~\cite{comanici2025gemini}) are included for context using results published
in~\cite{voxparadox}, without any intervention applied.

We apply IAAN to the audio encoder of each open-source model and of AF3-PD, and evaluate on VoxParadox~\cite{voxparadox}, a benchmark spanning ten non-semantic speech attributes (gender, age, emotion, intonation, speaker counting, speaker identity, speed, volume, vocal range, and pitch) with 200 multiple-choice question(MCQ) items per task. In each item, the voice carries one attribute while the spoken words imply a different one, referred to as the adversarial label, and both appear as answer options; correctly answering therefore requires genuine acoustic understanding rather than reading the transcript. Most tasks have four options (chance level 25\%), except gender which has two. We reproduce each model's baseline using its released inference code and evaluation protocol, with greedy decoding applied.

\subsection{Metrics}
\label{ssec:metrics}

We report \emph{accuracy} (Acc, \up{}) against ground truth and \emph{adversarial-label agreement} (ALA, \dn{}), the fraction of items where the model selects the adversarial label rather than the voice-carried attribute. A method that genuinely improves acoustic understanding should increase Acc and decrease ALA simultaneously: if ALA falls without Acc rising, predictions merely shift away from the adversarial label toward other wrong options rather than toward the acoustically grounded answer. We report per-task accuracy and average Acc/ALA for all baselines and \method{} variants. Sec.~\ref{ssec:hparam} uses a derived combination of these two quantities to select hyperparameters on a held-out development set.

\subsection{Hyperparameter Selection}
\label{ssec:hparam}
To keep VoxParadox strictly held out, we tune $(K,g)$ on an \emph{external} development set: the
$190$ sarcastic-speech clips from LISTEN~\cite{lexicalacoustic}, where the spoken emotion differs from the emotion the words imply. We rebuild each clip into an MCQ with both the voice emotion (ground truth) and the emotion implied by the words as options, where the latter serves as the adversarial label analogous to those in VoxParadox. For each model we choose the $K$ and $g$ that maximize the \emph{listening advantage} $\text{LA}=\text{Acc}-\text{ALA}$, which is higher when the model more often selects the voice-carried attribute and less often the adversarial label. We select on LA rather than raw accuracy because LA jointly rewards higher ground-truth accuracy and lower adversarial-label agreement, capturing the behavior we care about: predictions that are more acoustically grounded, and this signal may transfer more reliably to VoxParadox than raw accuracy alone. All reported \method{} results use the configurations chosen on the development set, not an oracle sweep over the test set.

\subsection{Compared Methods}
\label{ssec:controls}
In our main results (Table~\ref{tab:main}), we compare \method{} against three baselines that
intervene at other loci. Audio-aware decoding (AAD)~\cite{aad} contrasts output logits with and without audio at the
decoding side. LLM-amplify applies the same per-sample acoustic-rank criterion as \method{} but amplifies the same fraction of feed-forward neurons in the language model, operating exclusively at the positions of audio representations. HS-steer is a dense hidden-state steering baseline from~\cite{avs}, adapted to the encoder layer where most \method{}-selected neurons concentrate. Unlike the original method, which steers the language model's residual stream, our adaptation steers the encoder residual stream, contrasting dense representation-level steering with sparse neuron-level intervention at the same encoder locus.

%% file: sections/05_results_v2.tex
\section{Results}
\label{sec:results}
\input{tables/tab_main}

\subsection{Main Results}

Table~\ref{tab:main} reports the main VoxParadox results. Without intervention, the three open-source LALMs achieve only $7.6\%$--$21.2\%$ average accuracy, with average ALA reaching $64.0\%$--$76.9\%$. \method{} consistently improves accuracy while reducing ALA, raising accuracy by $25.7$, $21.4$, and $9.7$ points on AF3, Qwen2.5, and Kimi, respectively, while lowering ALA by $35.0$, $28.8$, and $21.0$ points. With \method{}, all three models surpass the closed-source systems in Table~\ref{tab:main} on both Acc and ALA, including Gemini 2.5 Flash (24.7\% Acc, 60.5\% ALA). This joint Acc increase and ALA decrease show that IAAN produces more acoustically grounded predictions rather than merely changing errors, and the gains hold across most attributes rather than a few.

We also apply \method{} to AF3-PD, a model explicitly fine-tuned to prioritize acoustic evidence~\cite{voxparadox}. \method{} still improves overall accuracy ($70.5\%\rightarrow75.0\%$) and reduces ALA ($22.4\%\rightarrow17.2\%$): on attributes where AF3-PD already performs well, accuracy is maintained or further improved, while on intonation, the attribute where AF3-PD still struggles, accuracy rises substantially ($15.5\%\rightarrow56.0\%$). This indicates that the encoder-side signal \method{} amplifies is complementary to what such training already supplies.

\subsection{Locus and Granularity Comparison}
\label{ssec:locus}
We compare \method{} with interventions at other model components to test whether the gains depend on acting inside the encoder and on selecting individual neurons rather than coarser directions (Table~\ref{tab:main}).

After the encoder, both audio-aware decoding (AAD), which contrasts
output logits, and LLM-amplify, which amplifies language-model
feed-forward neurons using the same acoustic-rank criterion as
\method{}, fail to recover the gain: AAD changes accuracy by at most a
few points in either direction across all four models. LLM-amplify
leaves AF3, Qwen2.5, and Kimi near baseline while reducing
AF3-PD substantially ($70.5\%\rightarrow40.2\%$). Neither contrasting logits nor amplifying neurons after the encoder recovers the acoustic signal that IAAN amplifies inside it.

At the same encoder locus as \method{}, hidden-state steering (HS-steer) adds a dense steering vector to the residual stream at the encoder layer where IAAN-selected neurons concentrate, typically near the encoder output (Sec.~\ref{ssec:neuron_position}). This contrasts dense representation-level steering with sparse neuron-level amplification. HS-steer yields only modest accuracy gains on AF3 and Qwen2.5, lowers accuracy on Kimi, and barely changes AF3-PD, reaching $20.9\%$, $13.7\%$, $18.2\%$, and $71.3\%$ accuracy for AF3, Qwen2.5, Kimi, and AF3-PD, respectively, compared with $38.8\%$, $29.0\%$, $30.9\%$, and $75.0\%$ for \method{}. Thus, acting at the right locus is not sufficient: a dense direction mixes the few neurons carrying acoustic evidence with many irrelevant dimensions, diluting the signal isolated by \method{}.

\method{} achieves the highest accuracy among all compared methods on every model while also lowering ALA, unlike LLM-amplify, whose lower ALA on AF3-PD ($22.4\%\rightarrow15.2\%$) comes with a collapse in accuracy ($70.5\%\rightarrow40.2\%$) rather than more voice-grounded answers. \method{} is the only method that consistently produces more acoustically grounded predictions across all four models.

\input{tables/tab_sweep}
\input{tables/tab_ref}

\subsection{Hyperparameter Sensitivity}

Table~\ref{tab:sweep} reports accuracy on VoxParadox and listening advantage on the LISTEN development set across the full $(K, g)$ grid. Hyperparameter sensitivity is model-dependent, ranging from AF3, which remains above the unamplified baseline across all configurations on both datasets, to Kimi, which has configurations falling below baseline on both. The optimal gain ranges from $g{=}8$ to $g{=}16$ across models, and the optimal neuron budget from $K{=}50$ to $K{=}200$, suggesting that too small a gain or too large a neuron budget is unlikely to be optimal, though the specific thresholds vary by model.

Despite this variation, nearly $90\%$ of configurations across all three models outperform the unamplified baseline on VoxParadox. This shows that the method does not require precise tuning: hyperparameter choice affects how much improvement is obtained, not whether improvement occurs at all.
The improvement found on the development set transfers to VoxParadox, with the configurations selected on the former losing within about 3 points relative to the per-model testing performance optimum, confirming that tuning on an external held-out set is sufficient. Developing principled methods for automatically selecting $(K, g)$ without a held-out development set remains an avenue for future work.

\subsection{Reference Ablation}
\label{ssec:reference_ablation}
\method{} contrasts each neuron's activation on the real clip with that on a reference (Eq.~\ref{eq:acoustic}); we ablate this
reference choice in Table~\ref{tab:ref}. Gaussian noise is a clip of
random noise constructed as described in Sec.~\ref{ssec:acoustic}, and
blank is a clip of silence. The noisy conditions instead add noise on
top of the original input clip at varying levels, with higher SNR
retaining a larger share of the original audio.

Gaussian noise and blank silence give similar performance across all
three models (Table~\ref{tab:ref}). At SNR\,$-20$\,dB, where the added noise dominates and little of the source clip remains audible, performance is also close to Gaussian noise and silence, as all three references carry little source-audio information.

As SNR increases and the reference preserves more of the source clip, accuracy drops on AF3 ($38.3\%$ at SNR\,$-20$\,dB to $21.9\%$ at SNR\,$+20$\,dB) and Qwen2.5 ($29.3\%$ to $10.8\%$), with ALA rising accordingly. Kimi is less sensitive, remaining within $28$--$31\%$ across SNRs. Even at high SNR, where the reference is less suitable for isolating acoustic sensitivity, \method{} remains above the unamplified baselines on AF3 and Qwen2.5 (Table~\ref{tab:ref}), suggesting that a suboptimal reference reduces but does not eliminate the gain.

This pattern is consistent with the contrast in Eq.~\ref{eq:acoustic}:
when the reference still carries some of the source clip's acoustic
content, that shared content is present in both passes and cancels out
of the difference, leaving a noisier estimate of which neurons respond
specifically to the real audio.

%% file: tables/tab_main.tex
\begin{table*}[t]
  \centering
  \caption{Accuracy (\%) and adversarial-label agreement (ALA, \%) on VoxParadox. $\dagger$: baseline reported in~\cite{voxparadox}; \emph{reproduced}: our unamplified baseline under the official protocol, i.e., the original model performance that all intervention methods build upon. \textbf{Bold}: best result per column within each model's block. Avg.\ and ALA report average accuracy and average adversarial-label agreement, respectively, across the ten tasks. Spk ID = speaker identity, Spk Cnt = speaker counting, Range = vocal range.}
  \vspace{-5pt}
  \label{tab:main}
  \small
  \setlength{\tabcolsep}{4pt}
  \renewcommand{\arraystretch}{1.2}
  \begin{tabular}{l cccccccccc | >{\columncolor{avgcol}}c >{\columncolor{alacol}}c}
    \toprule
    Model & Age & Gender & Emotion & Pitch & Volume & Speed & Range & Intonation & Spk ID & Spk Cnt & Avg.\,\up{} & ALA\,\dn{} \\
    \midrule
    \multicolumn{13}{l}{\textit{Open-source LALMs}} \\
    Audio-Flamingo-3$^\dagger$ & 10.0 & 16.0 & 24.5 & 11.0 & 11.5 & 11.5 & 9.5 & 34.0 & 23.5 & 22.5 & 17.4 & 68.5 \\
    \quad reproduced & 5.0 & 15.0 & 14.0 & 13.0 & 15.5 & 18.0 & 7.5 & 0.0 & 14.5 & 28.5 & 13.1 & 75.5 \\
    \quad AAD& 3.5 & 14.5 & 18.0 & 10.5 & 14.0 & 17.0 & 6.0 & 0.0 & 15.5 & 24.5 & 12.3 & 77.9 \\
    \quad HS-steer& 36.5 & 51.5 & 1.5 & 9.0 & 12.5 & 15.0 & 5.5 & 45.5 & 15.5 & 17.0 & 20.9 & 63.5 \\
    \quad LLM-amplify& 6.0 & 16.5 & 14.0 & 13.5 & 15.5 & 17.5 & 8.5 & 0.0 & 15.5 & 28.0 & 13.5 & 74.8 \\
    \quad \textbf{\method{} (Ours)} & \textbf{47.5} & \textbf{83.0} & \textbf{56.5} & \textbf{18.5} & \textbf{17.5} & \textbf{21.0} & \textbf{21.0} & \textbf{50.0} & \textbf{33.0} & \textbf{40.0} & \cellcolor{avghi}\textbf{38.8} & \cellcolor{alahi}\textbf{40.5} \\
    \midrule
    Qwen2.5-Omni$^\dagger$ & 1.5 & 2.0 & 3.0 & 13.0 & 8.0 & 8.0 & 12.5 & 4.5 & 15.5 & 11.5 & 8.0 & 75.2 \\
    \quad reproduced & 2.0 & 1.5 & 1.0 & 11.0 & 9.5 & 10.0 & 9.5 & 0.0 & 20.5 & 11.5 & 7.6 & 76.9 \\
    \quad AAD& 8.5 & 7.5 & 6.0 & 17.0 & 13.0 & \textbf{15.0} & 17.0 & 8.5 & 9.5 & 16.5 & 11.8 & 58.7 \\
    \quad HS-steer& 11.5 & 8.0 & 8.0 & 15.0 & 14.5 & 13.5 & 12.0 & 8.5 & \textbf{30.0} & 16.0 & 13.7 & 68.8 \\
    \quad LLM-amplify& 0.5 & 1.5 & 1.0 & 11.5 & 10.0 & 10.0 & 11.0 & 0.0 & 19.5 & 12.5 & 7.8 & 76.2 \\
    \quad \textbf{\method{} (Ours)} & \textbf{31.5} & \textbf{42.0} & \textbf{36.0} & \textbf{19.0} & \textbf{16.0} & \textbf{15.0} & \textbf{17.5} & \textbf{42.5} & 29.5 & \textbf{41.5} & \cellcolor{avghi}\textbf{29.0} & \cellcolor{alahi}\textbf{48.1} \\
    \midrule
    Kimi-Audio$^\dagger$ & 9.0 & 24.5 & 79.0 & 7.5 & 12.5 & 11.0 & 8.0 & 5.0 & 20.5 & 13.0 & 19.0 & 69.6 \\
    \quad reproduced & 12.0 & 33.5 & 83.5 & 9.5 & 19.0 & 12.5 & 13.0 & 5.5 & 13.0 & 10.5 & 21.2 & 64.0 \\
    \quad AAD& 12.5 & 31.0 & \textbf{84.5} & 9.5 & 18.0 & 11.5 & 11.5 & 5.5 & 14.0 & 8.5 & 20.6 & 63.6 \\
    \quad HS-steer& 13.0 & 21.0 & 74.0 & 10.5 & 16.0 & 13.0 & 11.0 & 4.5 & 10.0 & 9.5 & 18.2 & 68.0 \\
    \quad LLM-amplify& 12.0 & 34.0 & 83.5 & 9.5 & 18.0 & 13.5 & 13.5 & 6.0 & 13.5 & 10.0 & 21.4 & 63.6 \\
    \quad \textbf{\method{} (Ours)} & \textbf{30.5} & \textbf{59.0} & 18.0 & \textbf{19.5} & \textbf{25.0} & \textbf{18.5} & \textbf{21.5} & \textbf{39.5} & \textbf{32.5} & \textbf{45.0} & \cellcolor{avghi}\textbf{30.9} & \cellcolor{alahi}\textbf{43.0} \\
    \midrule
    \multicolumn{13}{l}{\textit{Specialized Tuned Model}} \\
    AF3-PD$^\dagger$ & 56.5 & 100.0 & 65.5 & 85.5 & 69.0 & 73.0 & 67.0 & 47.0 & 37.0 & 51.5 & 65.2 & 54.8 \\
    \quad reproduced & 77.0 & \textbf{100.0} & 47.0 & 98.5 & 96.5 & 90.0 & 96.5 & 15.5 & 27.5 & 56.5 & 70.5 & 22.4 \\
    \quad AAD& 77.5 & \textbf{100.0} & \textbf{48.5} & \textbf{99.0} & \textbf{99.0} & 90.5 & \textbf{97.0} & 12.0 & 27.0 & 57.5 & 70.8 & 22.1 \\
    \quad HS-steer& 76.5 & \textbf{100.0} & 45.5 & 98.5 & 97.0 & 90.5 & 96.5 & 21.5 & \textbf{31.0} & 56.0 & 71.3 & 21.2 \\
    \quad LLM-amplify& 50.5 & 40.5 & 25.0 & 70.5 & 52.5 & 50.0 & 60.0 & 12.5 & 20.5 & 20.0 & 40.2 & \cellcolor{alahi}\textbf{15.2} \\
    \quad \textbf{\method{} (Ours)} & \textbf{79.0} & \textbf{100.0} & 47.0 & 98.5 & 97.0 & \textbf{91.0} & 96.5 & \textbf{56.0} & 26.0 & \textbf{59.0} & \cellcolor{avghi}\textbf{75.0} & 17.2 \\
    \midrule
    \multicolumn{13}{l}{\textit{Proprietary Models}} \\
    GPT-4o Audio$^\dagger$ & 4.5 & 1.0 & 0.0 & 7.0 & 5.0 & 8.0 & 5.0 & 0.5 & 27.0 & 28.0 & 8.6 & 81.6 \\
    Gemini 2.5 Flash$^\dagger$ & 15.0 & 14.5 & 6.0 & 19.0 & 31.0 & 16.0 & 13.0 & 19.0 & 21.0 & 92.5 & 24.7 & 60.5 \\
    \bottomrule
  \end{tabular}
\end{table*}

%% file: tables/tab_sweep.tex
\begin{table}[t]

  \caption{Hyperparameter sweep over neuron budget $K$ and gain $g$. Left: LA${=}$Acc${-}$ALA (\%, $\uparrow$) on the LISTEN development set; right: Acc (\%, $\uparrow$) on VoxParadox. Unamplified baselines shown as development\,/\,test. \textbf{Bold}: best result per model per dataset; $\dagger$: configuration selected by maximizing development-set LA.}
  \vspace{-5pt}
  \label{tab:sweep}
  \scriptsize
  \setlength{\tabcolsep}{2pt}
  \renewcommand{\arraystretch}{1.2}
  \resizebox{0.99\columnwidth}{!}{%
  \begin{tabular}{@{}l >{\centering\arraybackslash}p{7.5mm} *{4}{>{\centering\arraybackslash}p{7.5mm}} *{4}{>{\centering\arraybackslash}p{7.5mm}}@{}}
    \toprule
    & & \multicolumn{4}{c}{\textit{LISTEN dev --- LA (\%)}} & \multicolumn{4}{c}{\textit{VoxParadox test --- Acc (\%)}} \\
    \cmidrule(lr){3-6}\cmidrule(lr){7-10}
    Model & $K$ & \multicolumn{4}{c}{gain $g$} & \multicolumn{4}{c}{gain $g$} \\
    & & 4 & 8 & 12 & 16 & 4 & 8 & 12 & 16 \\
    \midrule
    \multirow{4}{*}{\shortstack[l]{Audio-Flamingo-3\\[2pt]\scriptsize \textit{Baseline (unamplified):} \\27.4 / 13.1}}
        & 50   & 47.4 & 54.2 & 55.3 & \textbf{60.0} & 31.9 & \textbf{41.5} & 40.8 & 38.8$^\dagger$ \\
        & 100  & 46.8 & 53.2 & 54.2 & 54.2 & 30.3 & 38.5 & 36.9 & 34.5 \\
        & 200  & 49.5 & 53.7 & 57.4 & 53.7 & 30.8 & 35.0 & 32.5 & 31.2 \\
        & 1000 & 49.0 & 45.8 & 49.0 & 32.6 & 31.1 & 32.9 & 31.6 & 30.6 \\
    \midrule
    \multirow{4}{*}{\shortstack[l]{Qwen2.5-Omni\\[2pt]\scriptsize \textit{Baseline (unamplified):} \\15.3 / 7.6}}
        & 50   & 18.4 & 34.7 & \textbf{35.8} & 31.1 & 8.6 & 11.9 & 29.0$^\dagger$ & 32.2 \\
        & 100  & 13.7 & 33.7 & 33.2 & 33.2 & 8.9 & 18.9 & \textbf{30.5} & 30.4 \\
        & 200  & 15.8 & 35.3 & 30.5 & 26.3 & 8.3 & 23.9 & 30.1 & 29.4 \\
        & 1000 & 22.1 & 35.3 & 32.1 & 27.4 & 7.8 & 29.2 & 29.4 & 29.5 \\
    \midrule
    \multirow{4}{*}{\shortstack[l]{Kimi-Audio\\[2pt]\scriptsize \textit{Baseline (unamplified):} \\-14.7 / 21.2}}
        & 50   & $-$13.7 & $-$13.7 & $-$1.6 & $-$9.5 & 21.3 & 20.6 & 21.3 & 24.6 \\
        & 100  & $-$21.1 & $-$6.3 & $-$7.9 & $-$4.2 & 21.1 & 15.1 & 29.9 & 28.4 \\
        & 200  & $-$21.6 & 0.0 & \textbf{6.8} & $-$6.8 & 20.3 & 27.8 & \textbf{30.9}$^\dagger$ & 29.2 \\
        & 1000 & $-$14.2 & $-$9.5 & $-$14.7 & $-$36.3 & 11.8 & 27.7 & 27.8 & 27.2 \\
    \bottomrule
  \end{tabular}%
}
\end{table}

%% file: tables/tab_ref.tex
\begin{table}[ht]
  \centering
  \caption{Reference ablation on VoxParadox (Acc and ALA \%). We compare performance under different reference signal choices. The unamplified baseline is included for comparison. \textbf{Bold}/\underline{underline}: best/second-best per column.}
  \vspace{-5pt}
  \label{tab:ref}
  \small
  \setlength{\tabcolsep}{2pt}
  \renewcommand{\arraystretch}{1.2}
  \resizebox{\columnwidth}{!}{
  \begin{tabular}{l cc cc cc}
    \toprule
    Model & \multicolumn{2}{c}{Audio-Flamingo-3} & \multicolumn{2}{c}{Qwen2.5-Omni} & \multicolumn{2}{c}{Kimi-Audio} \\
    \cmidrule(lr){2-3} \cmidrule(lr){4-5} \cmidrule(lr){6-7}
    Reference Type & Acc\,\up{} & ALA\,\dn{} & Acc\,\up{} & ALA\,\dn{} & Acc\,\up{} & ALA\,\dn{} \\
    \midrule
    \textit{Baseline (unamplified)} & 13.1 & 75.5 & 7.6 & 76.9 & 21.2 & 64.0 \\
    \midrule
    Gaussian noise & \underline{38.8} & 40.5 & \underline{29.0} & \textbf{48.1} & \textbf{30.9} & 43.0 \\
    Blank (silence)   & \textbf{39.4} & \textbf{38.1} & 28.1 & \underline{49.2} & 28.1 & \textbf{42.3} \\
    Noisy SNR\,$-20$ dB  & 38.3 & \underline{39.1} & \textbf{29.3} & \textbf{48.1} & 29.7 & \underline{42.7} \\
    Noisy SNR\,$-10$ dB  & 29.1 & 50.1 & 24.6 & 49.6 & 28.4 & 43.4 \\
    Noisy SNR\,$0$ dB    & 19.5 & 63.7 & 14.3 & 68.1 & 28.5 & 43.9 \\
    Noisy SNR\,$+10$ dB  & 18.3 & 65.9 & 12.4 & 72.7 & 28.1 & 44.7 \\
    Noisy SNR\,$+20$ dB  & 21.9 & 62.7 & 10.8 & 73.6 & \underline{30.6} & 43.7 \\
    \bottomrule
  \end{tabular}
  }
\end{table}

%% file: sections/06_analysis_v2.tex
\input{figs/fig_acoustic_v2}
\section{Analysis}
\label{sec:analysis_v2}

\subsection{Acoustic Score Distribution and Neuron Localization}
\input{tables/tab_layersteer}
\label{ssec:neuron_position}
\method{} amplifies at most $0.12\%$ of the encoder's feed-forward neurons ($K{=}200$ out of $N{=}163{,}840$), yet raises accuracy substantially, by $+9.7$ to $+25.7$ points. Such a large effect from so few neurons leads us to examine the distribution of the acoustic score itself, to see whether responsiveness to acoustic information is broadly shared across the encoder or concentrated in a small subset of neurons. Taking AF3 as an example, Fig.~\ref{fig:acoustic_hist} shows that the acoustic score is sharply concentrated near zero: the great majority of neurons respond almost identically to the real audio and to the noise reference. Only a small tail breaks away from this majority. The $K{=}50$ selection threshold, $s_i{=}0.40$, sits about $13$ standard deviations above the mean of this score distribution, well outside the majority rather than at an arbitrary cut along a smooth distribution.

Fig.~\ref{fig:acoustic_layers} indicates the selected neurons are unevenly distributed across encoder layers. For AF3 and Qwen2.5, the majority of them fall in the encoder's last layer, whereas for Kimi they are spread more widely across layers; even so, the final layer still holds a higher share of selected neurons than any other individual layer, so the tendency toward the encoder's output holds to some degree across all three models. Why later layers exhibit this stronger acoustic-discriminative response remains an open question; further investigation could shed light on how acoustic information is organized across the encoder's depth, which we leave to future work.

\subsection{Selection of the Acoustic Neurons Drives the Gain}
\label{ssec:selection}

\input{tables/tab_qual}

Sec.~\ref{ssec:neuron_position} shows that the neurons \method{} selects are often concentrated
in a single encoder layer. This raises the question of whether the gain depends on selecting
these particular neurons, or whether amplifying that layer as a whole would work just as well.
We test this with three controls on the encoder, each isolating one factor: random selection
(Random), whole-layer amplification (Layer-amplify), and selection restricted to the layer
where most of \method{}'s selected neurons concentrate (IAAN (Layer-restricted)). Since Layer-amplify operates at a much larger scale than a small neuron subset, we re-select its gain $g$
separately on the same LISTEN development set (Sec.~\ref{ssec:hparam}).
Table~\ref{tab:layersteer} reports VoxParadox performance for all three controls. We focus the
discussion below on accuracy; ALA follows the same trends across all three controls.

Random selection keeps the neuron budget fixed but replaces \method{}'s selected neurons with the same number of randomly chosen encoder neurons, averaged over three random seeds.
 It offers at small
improvement over baseline across all three models, far short of \method{}'s gains, indicating
that the gain comes from the acoustic-score selection itself, not merely from amplifying the
same number of neurons. Layer-amplify removes selection entirely, uniformly scaling every
neuron in the layer where most of \method{}'s selected neurons concentrate, at its best gain. It
stays at baseline, indicating that amplifying this layer without selecting within it does not
help either.

We then evaluate IAAN (Layer-restricted): for each model, we identify the encoder layer containing the largest share of neurons selected by full \method{}, then rerun the selection within that layer using the same neuron budget. This tests whether \method{} still works once the layer is fixed.
For Qwen2.5, this recovers the gain in
full, reaching the same accuracy as full \method{} ($29.0\%$); for AF3, it recovers
most of the gain ($37.2\%$ against $38.8\%$). On Kimi, however, this restriction performs
below the unamplified baseline ($20.7\%$ against $21.2\%$) and falls well short of full
\method{} ($30.9\%$). Unlike AF3 and Qwen2.5, Kimi's selected neurons
are not as strongly concentrated in a single layer, and a considerable portion lie outside
whichever layer the restriction targets. Restricting the search to a single layer therefore
excludes many of the neurons \method{} would otherwise select. This indicates that searching
across the full encoder, as \method{} does, is necessary to reach these neurons on models like
Kimi, where their distribution across layers is less concentrated.

Combining these controls with the locus and granularity comparison (Sec.~\ref{ssec:locus}), we confirm that both the encoder as the intervention site and the selectivity of the neuron set within it are necessary for \method{}'s gain. These results further suggest that the acoustic score provides a reliable signal for identifying which neurons matter across the full encoder, even without task labels or a calibration set.

\subsection{Qualitative Results}
\label{sec:qual}

Beyond multiple-choice evaluation, we examine free-text descriptions to test whether \method{} changes how the model describes acoustic properties, rather than only which option it selects (Table~\ref{tab:qual}). Across the three examples, the baseline often misdescribes key cues such as gender, emotion, and speaker count, while \method{} produces descriptions more consistent with the actual voice. These shifts mirror the quantitative results: amplifying acoustic neurons helps the model ground its descriptions in how the audio sounds.

%% file: figs/fig_acoustic_v2.tex
\captionsetup[subfigure]{skip=0pt}
\begin{figure}[t]
    \centering
    \subfloat[Acoustic-score distribution\label{fig:acoustic_hist}]{
        \includegraphics[width=\linewidth]{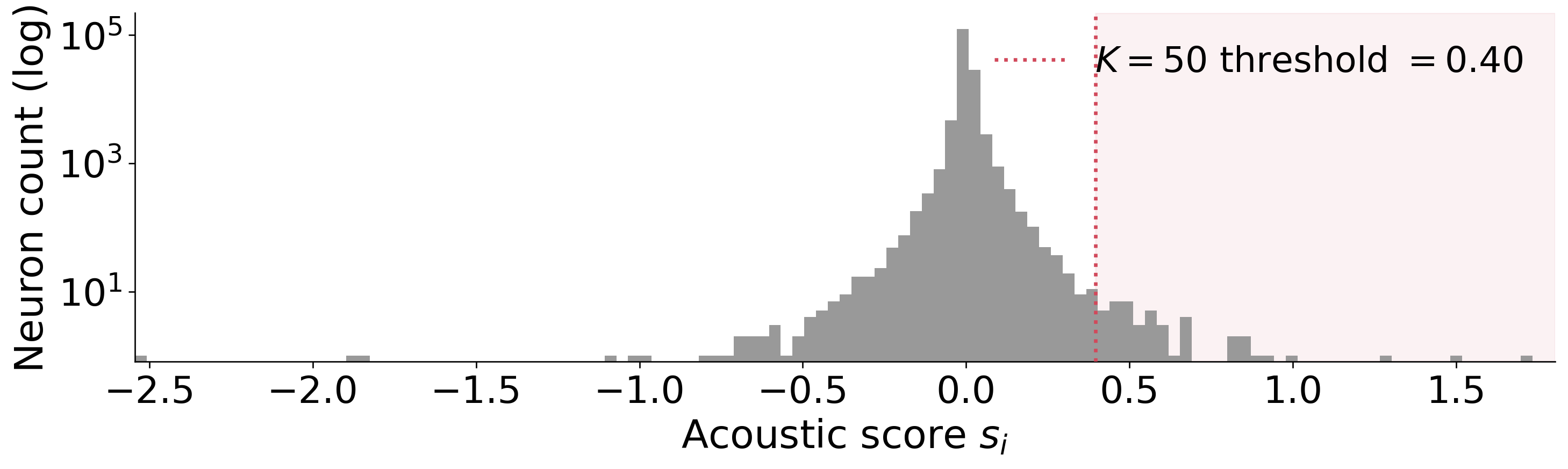}
    }\\[6pt]
    \subfloat[Selected neurons per encoder layer\label{fig:acoustic_layers}]{
        \includegraphics[width=\linewidth]{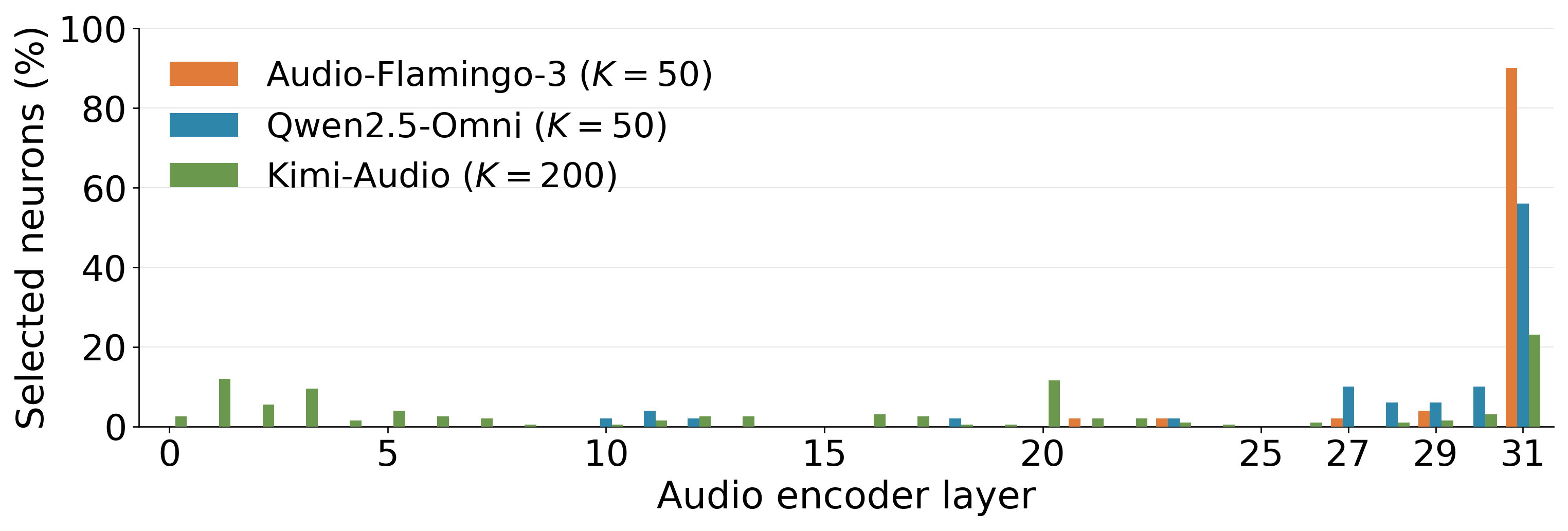}
    }
    \caption{Acoustic-score distribution and layer localization. Statistics are averaged over all test examples. (a) Distribution of the acoustic score $s_i$ over AF3 neurons; the red dashed line marks the selection threshold. (b) Share of selected neurons in each audio encoder layer.}
    \label{fig:acoustic}
\end{figure}

%% file: tables/tab_layersteer.tex
\begin{table}[ht]
  \centering
  \caption{VoxParadox Acc and ALA (\%) for controls on neuron selection. \textbf{Bold}: best per column.}
  \vspace{-5pt}
  \label{tab:layersteer}
  \small
  \setlength{\tabcolsep}{3pt}
  \renewcommand{\arraystretch}{1.2}
  \resizebox{\columnwidth}{!}{%
  \begin{tabular}{l cc cc cc}
    \toprule
    Model & \multicolumn{2}{c}{Audio-Flamingo-3} & \multicolumn{2}{c}{Qwen2.5-Omni} & \multicolumn{2}{c}{Kimi-Audio} \\
    \cmidrule(lr){2-3}\cmidrule(lr){4-5}\cmidrule(lr){6-7}
    Method & Acc\,\up{} & ALA\,\dn{} & Acc\,\up{} & ALA\,\dn{} & Acc\,\up{} & ALA\,\dn{} \\
    \midrule
    \textit{Baseline (unamplified)} & 13.1 & 75.5 & 7.6  & 76.9 & 21.2 & 64.0 \\
    \midrule
    Random                & 13.2 & 75.3 & 7.8  & 76.6 & 24.6 & 60.3 \\
    Layer-amplify         & 12.4 & 76.7 & 7.8  & 77.0 & 21.5 & 63.9 \\
    \method{} (Layer-restricted) & 37.2 & 41.1 & 29.0 & 48.1 & 20.7 & 62.8 \\
    \midrule
    \method{}             & \textbf{38.8} & \textbf{40.5} & \textbf{29.0} & \textbf{48.1} & \textbf{30.9} & \textbf{43.0} \\
    \bottomrule
  \end{tabular}}
\end{table}

%% file: tables/tab_qual.tex
\begin{table*}[ht]
  \centering
  \caption{Qualitative examples showing how \method{} changes free-text acoustic descriptions on AF3. For each attribute, the Input column gives the spoken content and two labels: the \qgt{voice attribute} (ground truth) and the \qadv{adversarial label}.}
  \label{tab:qual}
  \footnotesize
  \setlength{\tabcolsep}{4pt}
  \renewcommand{\arraystretch}{1.0}

  \begin{tabularx}{\linewidth}{
    @{}
    >{\raggedright\arraybackslash}m{0.07\linewidth}
    >{\raggedright\arraybackslash}m{0.25\linewidth}
    >{\centering\arraybackslash}m{0.09\linewidth}
    Y
    @{}
  }
  \toprule
  \textbf{Attribute}
  & \textbf{Input}
  & \textbf{Method}
  & \textbf{Output} \\
  \midrule

  \multirow{2}{=}{\textbf{Gender}}
  & \multirow{2}{=}{\inputcell
    {As a female, I feel empowered to share my story.}
    {male}
    {female}}
  & \makecell[c]{\textit{Baseline}\\\textit{(unamplified)}}
  & \qx~``The speaker is a young adult \qadv{female} with a high-pitched, confident, and assertive voice. She sounds empowered and self-assured, with a clear and steady tone that conveys strength and determination.'' \\
  \cmidrule(l){3-4}
  &
  &
  \method{}
  & \qck~``The speaker is a young \qgt{male} with a high-pitched, angry voice, speaking in a flat, monotone manner.'' \\
  
  \midrule

  \multirow{2}{=}{\textbf{Emotion}}
  & \multirow{2}{=}{\inputcell
    {Right now, my heart is calm and steady.}
    {angry}
    {neutral}}
  & \makecell[c]{\textit{Baseline}\\\textit{(unamplified)}}
  & \qx~``The speaker in this audio is a female with a \qadv{calm and steady} voice.'' \\
  \cmidrule(l){3-4}
  &
  &
  \method{}
  & \qck~``The speaker in this audio is a young male with a high-pitched, slightly nasal voice. He sounds \qgt{slightly annoyed and impatient}, and his voice is quite monotone.'' \\

 \addlinespace[2pt]
  \midrule

  \multirow{2}{=}{\textbf{Speaker count}}
  & \multirow{2}{=}{\inputcell
    {I am alone in this discussion. Yes, absolutely. Just one person.}
    {two speakers}
    {one}}
  & \makecell[c]{\textit{Baseline}\\\textit{(unamplified)}}
  & \qx~``I hear \qadv{one person} speaking in this audio. The speaker is a male with a deep voice.'' \\
  \cmidrule(l){3-4}
  &
  &
  \method{}
  & \qck~``I hear \qgt{two distinct speakers} in this audio. One is a male voice speaking in a deep, resonant tone, and the other is a female voice speaking in a higher, more melodic tone.'' \\

  \bottomrule
  \end{tabularx}
\end{table*}

%% file: sections/08_conclusion.tex
\section{Conclusion}
\label{sec:conclusion}

We present IAAN, a training-free and label-free method that improves acoustic understanding in large audio-language models by intervening inside the audio encoder. IAAN scores each neuron by how much more it responds to the real waveform than to a noise reference lacking the real audio's acoustic information, then amplifies a small set of the highest-scoring neurons at inference. Across three open-source LALMs, IAAN modifies at most 0.12\% of the encoder's neurons, yet improves average accuracy by 9.7 to 25.7 points, with gains broad across attributes rather than concentrated in a few. Controlled experiments further confirm that both the encoder locus and the selectivity of the neuron set are essential to this gain. These findings establish the audio encoder's individual neurons as an effective and largely untapped site for inference-time intervention, suggesting that deeper exploration of the encoder's internal structure may yield further gains in acoustic understanding.